\documentclass[conference]{IEEEtran}
\IEEEoverridecommandlockouts
\usepackage{cite}
\usepackage{amsmath,amssymb,amsfonts}
\usepackage{algorithmic}
\usepackage{graphicx}
\usepackage{multirow}
\usepackage{verbatim}
\usepackage{textcomp}
\usepackage{xcolor}
\usepackage{makecell}
\usepackage{bm}
\usepackage{graphicx}
\usepackage{subfigure}

\usepackage[ruled,linesnumbered]{algorithm2e}

\def\BibTeX{{\rm B\kern-.05em{\sc i\kern-.025em b}\kern-.08em
    T\kern-.1667em\lower.7ex\hbox{E}\kern-.125emX}}
\begin{document}

\title{Wireless Transmission of Images with the Assistance of Multi-level Semantic Information\\
\thanks{This work is partly supported by National Key R\&D Program of China under the project 2020YFB1708700, partly by the SUTD-ZJU IDEA Grant (SUTD-ZJU (VP) 202102), and partly by the Fundamental Research Funds for the Central Universities under Grant 2021FZZX001-20.}
}

\author{\IEEEauthorblockN{Zhenguo Zhang, Qianqian Yang, Shibo He, Mingyang Sun, Jiming Chen}
\IEEEauthorblockA{{
}
{Zhejiang University, Zhejiang, China}\\
$\{\rm{zhangzhenguo, qianqianyang20, s18he, mingyangsun, 
  cjm}\}$@zju.edu.cn}

}
\maketitle

\begin{abstract}
Semantic-oriented communication has been considered a promising method to boost bandwidth efficiency by only transmitting the semantics of the data. In this paper, we propose a multi-level semantic aware communication system for wireless image transmission, named MLSC-image, which is based on deep learning (DL) techniques and trained in an end-to-end manner. In particular, the proposed model includes a multi-level semantic feature extractor, that extracts both the high-level semantic information, such as the text semantics and the segmentation semantics, and the low-level semantic information, such as local spatial details of the images. We employ a pre-trained image caption to capture the text semantics and a pre-trained image segmentation model to obtain the segmentation semantics. These high-level and low-level semantic features are then combined and encoded by a joint semantic and channel encoder into symbols to transmit over the physical channel. The numerical results validate the effectiveness and efficiency of the proposed semantic communication system, especially under the limited bandwidth condition, which indicates the advantages of the high-level semantics in the compression of images. 


\end{abstract}

\section{Introduction}
Shannon’s separation theorem is the cornerstone of the traditional communication system\cite{shannon1948mathematical}, serving as a design criterion to establish modern wireless communication framework from the first-generation (1G) to the fifth-generation (5G). With the rapid development of wireless communication, the communication system channel capacity is close to the Shannon limit. Researchers have been exploring new ways to meet the rapid-growing demands of wireless data transmission, and the semantic oriented communication system has been well recognized as a promising approach for the next generation of wireless communication\cite{strinati20216g,strinati2021toward,chengxiang46g}.

In contrast to traditional communication, semantic communication focuses on transmitting the meaning of the information, instead of the exact transmission of symbols. The existing semantic communication systems jointly design the transmitter and receiver to achieve better transmission efficiency and robustness to the varying channel conditions\cite{shi2021semantic,basu2014preserving}. 
DL-based semantic communication systems have revealed tremendous potential in the efficient transmission of different types of information, i.e., text\cite{xie2020lite,farsad2018deep,xie2021deep}, speech\cite{weng2021semantic,tong2021federated}, and image\cite{bourtsoulatze2019deep,qiu2020deep,kurka2020deepjscc,yang2021deep,lee2019deep,jankowski2020wireless,jankowski2020joint}.

For the semantic compression and transmission of the images, DL has been applied to conventional image compression\cite{qiu2020deep}, which proposes a two-step method by combining the state-of-the-art signal processing-based recovery method with a deep residual learning model to recover the original image. Bourtsoulatze \textit{et al}. \cite{bourtsoulatze2019deep} propose a DL-based wireless image transmission system to achieve joint source-channel coding in an end-to-end manner (JSCC), where peak signal-to-noise ratio (PSNR) and structure-similarity-index-measure (SSIM) are devoted to measuring the quality of the reconstructed images. Based on JSCC, another image reconstruction scheme with channel feedback, named DeepJSCC-f, has been implemented in \cite{kurka2020deepjscc} to enhance image reconstruction accuracy with channel feedback from the receiver. Yang \textit{et al}. \cite{yang2021deep} present a JSCC scheme with an OFDM datapath for wireless image transmission over multipath fading channels. Moreover, the proposed model achieves remarkable performance by integrating expert knowledge. The progressive research on Internet-of-Things (IoT) devices for image transmission was developed in\cite{lee2019deep}, which proposes a joint transmission-recognition scheme leveraging two deep neural network (DNNs) and outperforming the conventional scheme on the IoT devices. Jankowski \textit{et al}. \cite{jankowski2020wireless} present a joint feature compression and transmission system to deal with the limited computational resources at the edge server. The scheme not only improves end-to-end reliability but also reduces computation complexity. Jankowski \textit{et al}. \cite{jankowski2020joint} proposes an autoencoder-based system for device-edge communication with rigorous constraints, which achieves better performance in classification accuracy with limited computation capacity.

In this work, we propose a novel DL-based semantic communication system for image transmission over noisy wireless channels. In particular, we introduce a multi-level semantic feature extractor, that exploits both the high-level semantic information including the text semantic information and segmentation semantic information, which contains abstractedness and generality indication of an image\cite{liu2020adaptive}, and the low-level semantic information, such as local details of the image. The numerical results show that the proposed method remarkably outperforms the existing method in a high to medium compression ratio, which validates the semantic significance of the high-level semantic information.




The rest of the paper is organized as follows. In Section \textrm{II}, we introduce the system model. The details of the architecture are shown in \textrm{III}. Section \textrm{IV} presents the numerical results, and Section \textrm{V} concludes the paper.

\section{System model}

\subsection{System Description}

We consider a semantic communication system, as shown in Fig. 1, where the transmitter sends an image to the receiver over a physical channel. The system consists of two parts: (\textit{i}) the encoder at the transmitter, which extracts semantic features from the input image and encodes them to symbols to transmit over the channel; and (\textit{ii}) the decoder at the receiver, which decodes the semantic features from the received signals to reconstruct the input image from them. We will describe the encoder and decoder in detail in the sequel.

\begin{figure}[t]
\centering
\includegraphics[scale=0.82]{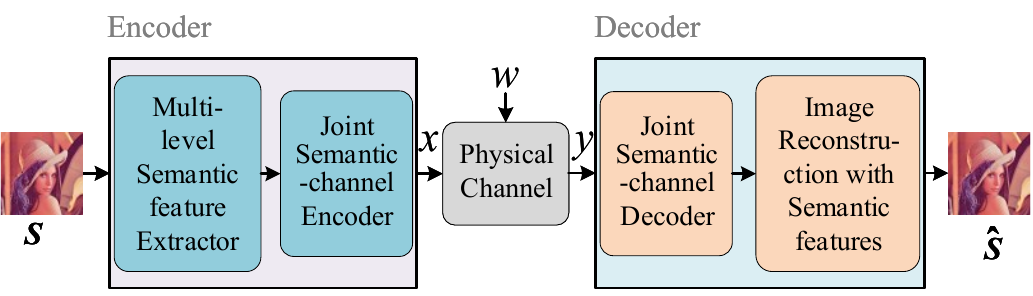}
\caption{The proposed semantic-oriented communication of image.}
\label{fig:1}
\end{figure}

\subsection{Encoder}
The encoder consists of two components: the multi-level semantic feature extractor and the joint semantic-channel encoder. The input image $\bm{S}$ to the encoder is preprocessed by a normalization layer such that each element is in the range of [0, 1]. Then the multi-level semantic feature extractor, which contains multiple NN-based modules, extracts different semantic features of the input image. The joint semantic-channel encoder encodes these semantic features into symbols to transmit over the physical channel to the receiver. Denote the NN parameters of the multi-level semantic feature extractor and the joint semantic-channel encoder as $\alpha$, and $\beta$, respectively. Then the signal symbol vector to be transmitted $\bm{x}$ is given:

\begin{equation}
\bm{x}=\textbf{T}_\beta(\textbf{T}_\alpha(\bm{S})), 
\end{equation}

\noindent where $\textbf{T}_\alpha(\cdot)$ denotes the operation by the multi-level features extracting network with parameters $\alpha$, and $\textbf{T}_\beta(\cdot)$ is the joint semantic-channel encoder function with parameters $\beta$.




We consider a widely used physical channel: the additive white Gaussian noise (AWGN) channel. We have the received signal $\bm{y}$ at the decoder given as:

\begin{equation}
\bm{y}=\bm{x}+\bm{w},
\end{equation}

\noindent where $\bm{w}$ is a vector of independent identically distributed (i.i.d.) channel noise following a circularly symmetric Gaussian distribution, i.e., $\bm{w}\sim\mathcal{CN}(0,\sigma^2\textbf{I})$, $\sigma^2$ is the average noise power of the channel, and $\textbf{I}$ is the identity matrix. 

\begin{figure*}[t]
\centering
\includegraphics[scale=0.67]{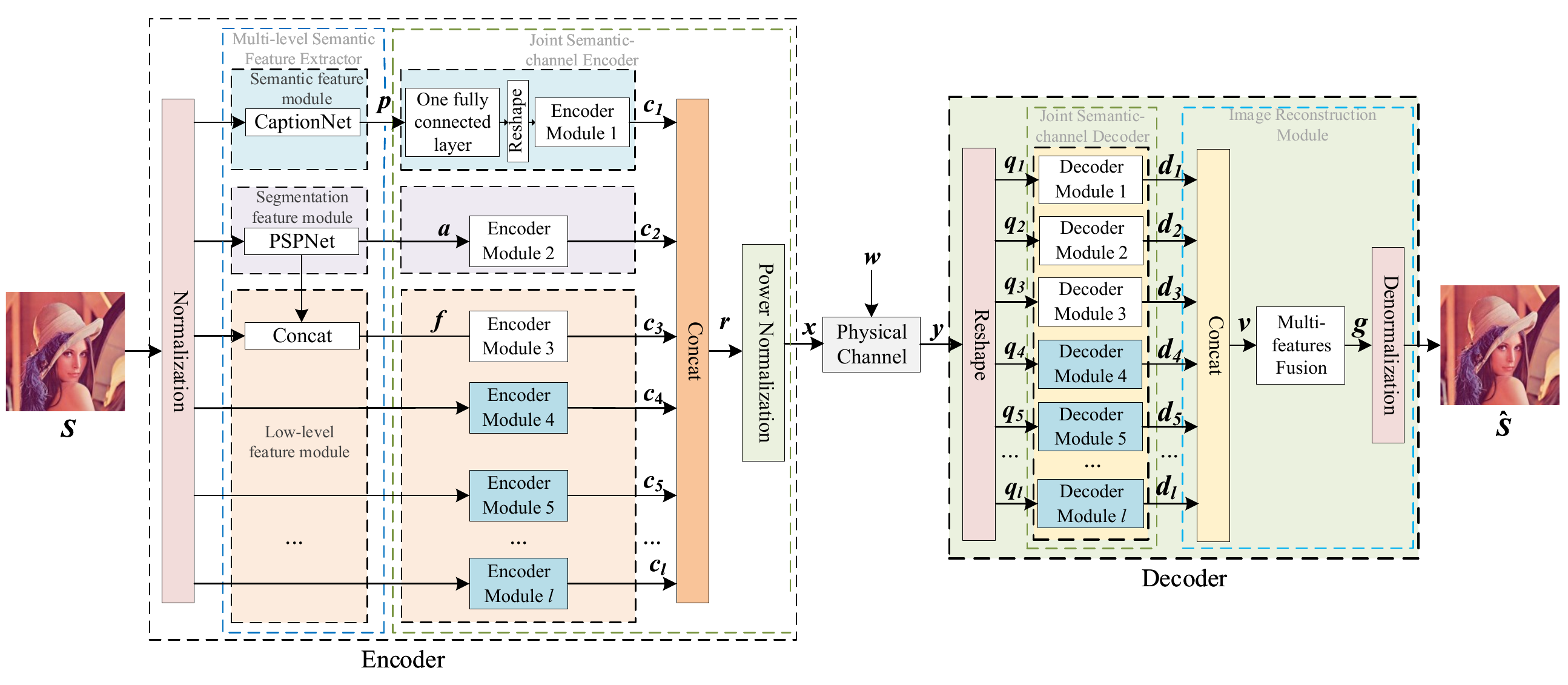}
\caption{The overall architecture of the proposed semantic communication system, MLSC-image.}
\label{fig:2}
\end{figure*}

\begin{figure}[!hb]
\centering
\includegraphics[scale=0.38]{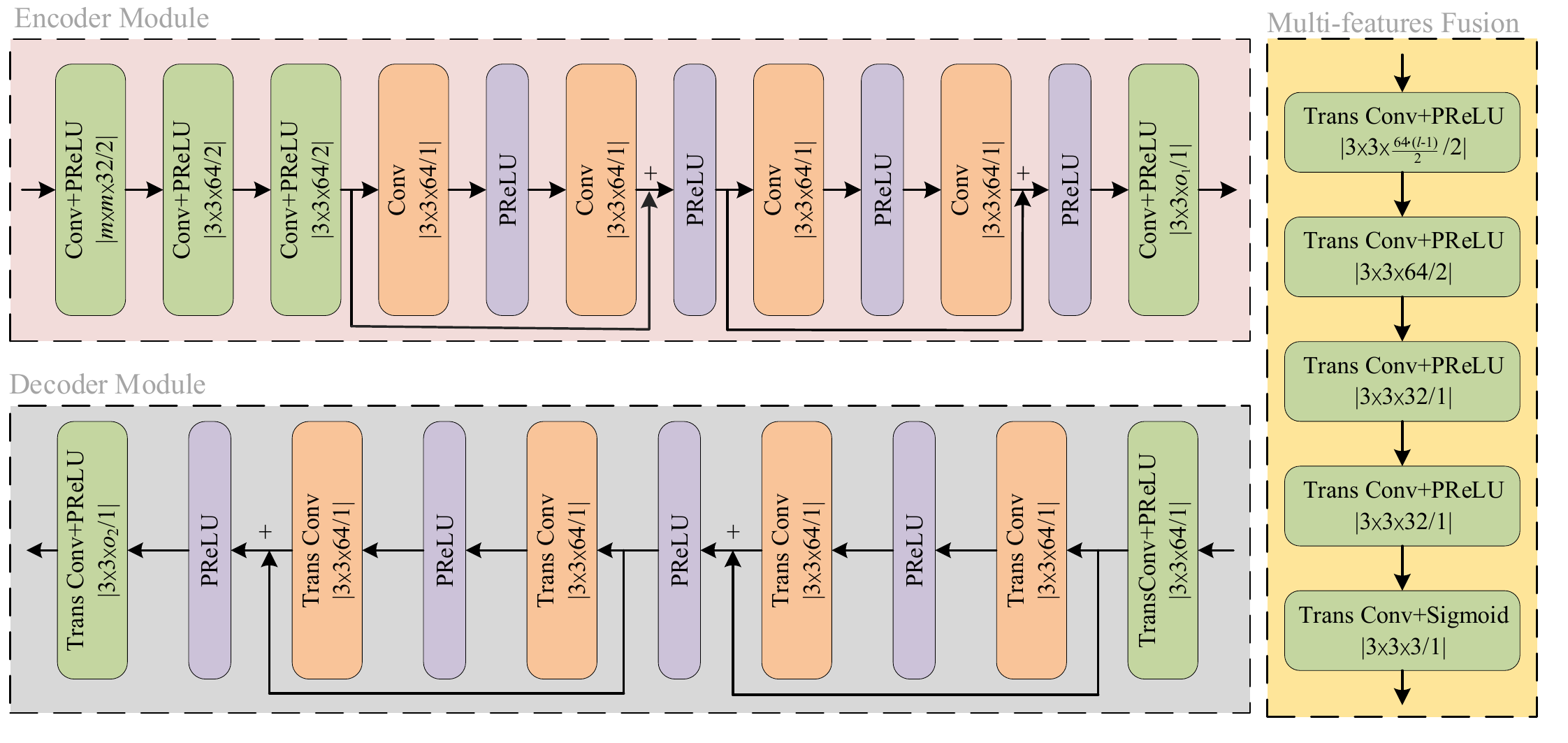}
\caption{The architecture of the Encoder Module, the Decoder Module and the Multi-features Fusion Module, the convolutional neural networks are parameterized by $m\times{m}\times{o/st}$, where \textit{m, o} and \textit{st} are kernel size, the number of channel outputs, and the stride, respectively. Each convolutional layer is followed by a generalized normalization transformation layer (GDN/IGDN), which we omit here due to the space limitation.}
\label{fig:3}
\end{figure}

\subsection{Decoder}
The decoder consists of two parts, the joint semantic-channel decoder, and the image reconstruction module. The joint semantic-channel decoder mitigates physical noise in the received signal collected over the AWGN channel and recovers the multi-level semantic features. The image reconstruction module fuses the different levels of semantic information and reconstructs the target image. Finally, the denormalization layer rescales each element to image pixel values (0-255).

The parameter sets of the joint semantic-channel decoder and the image reconstruction module are denoted by $\zeta$ and $\eta$, respectively. The reconstructed image at the receiver from the received signal is given by:
\begin{equation}
\hat{\bm{S}}=\textbf{R}_\eta(\textbf{R}_\zeta(\bm{y})),
\end{equation}

\noindent where $\textbf{R}_\zeta(\cdot)$ and $\textbf{R}_\eta(\cdot)$ denote the operation of the joint semantic-channel decoder and the image reconstruction module, respectively. 

The goal of the considered semantic communication system is to minimize the average distortion between input image $\bm{S}$ and reconstructed image $\hat{\bm{S}}$. The mean-squared-error (MSE) is used as the loss function to evaluate the distortion between $\bm{S}$ and $\hat{\bm{S}}$, denoted as:

\begin{equation}
\mathcal{L}=\frac{1}{N}	\sum_{k=1}^N d(\bm{S}_k,\hat{\bm{S}}_k),
\end{equation}

\noindent where $d(\bm{S},\hat{\bm{S}})=\frac{1}{n}||\bm{S}-\hat{\bm{S}}||^2$ is the mean squared-error distribution, and \textit{N} is the number of samples.


\section{Proposed Method}
We present the proposed DL-based semantic communication system for wireless image transmission, as shown in Fig. 2, referred to as MLSC-image. Specifically, the multi-level semantic feature extractor is used for extracting different levels of semantic features. The joint semantic-channel encoder and decoder successfully recover these features at the receiver. Then the image reconstruction module is adopted to fuse the multi-level semantic information so that to reconstruct the target image. 

\subsection{Multi-level Semantic Feature Extractor}
The input of the proposed MLSC-image scheme, denoted as $\bm{S}\in\mathfrak{R}^{b\times{h}\times{w}\times{3}}$, is a batch of image samples, where \textit{b, h}, and \textit{w} are the batch size, the image height, and width, and the last dimension corresponds to the RGB channels. The image is first preprocessed by a normalization layer to scale the value of each pixel to [0, 1]. A batch of normalized images is then input to the multi-level semantic feature extractor, which consists of three-level feature extraction modules: the semantic feature module, the segmentation feature module, and the low-level feature module. The semantic feature module contains a pre-trained image caption model, which is frozen during the training of the whole systemm, to acquire text-form semantic features $\bm{p}\in\mathfrak{R}^{\frac{b\times2\cdot{h}\cdot{w}}{t^2}\times{1}}$, where \textit{t} is the downsampling factor, which consists of a ResNet-152 model \cite{he2016deep}and a LSTM layer \cite{yu2019review}. The text-form semantic features are the text embeddings with image texture information, and we call the features ``text-form" for understanding. The segmentation feature module uses a pre-trained semantic image segmentation network, PSPNet \cite{zhao2017pyramid} to obtain the high-level segmentation features $\bm{a}\in\mathfrak{R}^{b\times{h}\times{w}\times{1}}$. Then, the low-level feature module consists of \textit{l}-2 components: the Concat part, which concatenates the normalized image and the segmentation result, and outputs $\bm{f}\in\mathfrak{R}^{b\times{h}\times{w}\times{4}}$; and \textit{l}-3 repeated direct input of the normalized image. This module aims to extract low-level semantic features, such as local details of the images. The hyperparameter \textit{l} governs how many complementary details of the image are to transmit to the receiver. This controls the trade-off between the quality of the reconstructed image and the transmission cost. The larger the \textit{l} is, the better reconstruction will be, and vice versa. These semantic features are then input to the semantic-channel encoder to be encoded into transmitted symbols.

\subsection{Joint Semantic-channel Encoder and Decoder}

The joint semantic-channel encoder takes in four types of input: the text-form features $\bm{p}$, the semantic image segmentation $\bm{a}$, the concatenated features $\bm{f}$, and the normalized image. The semantic features $\bm{p}$ is fed into a fully connected layer to reduce the dimension, followed by a reshape layer and encoder module, which outputs $\bm{c}_1\in\mathfrak{R}^{b\times{\frac{h}{t}\times{\frac{w}{t}}}\times{1}}$.
The semantic features $\bm{a}$ and $\bm{f}$ are fed into two encoder modules, respectively. The structure of the module is presented in Fig. 3. These two encoder modules generate outputs of different dimensions, i.e., $\bm{c}_2\in\mathfrak{R}^{b\times{\frac{h}{t}\times{\frac{w}{t}}}\times{1}}$ and $\bm{c}_3\in\mathfrak{R}^{b\times{\frac{h}{t}\times{\frac{w}{t}}}\times{3}}$. The encoder module \textit{i} ($i=4, \ldots, l$) shares the same model shown in Fig. 3 with various kernel sizes $\bm{m}=2\times{i}-5$. Each of these modules takes the normalized image directly as the input, and output $\bm{c}_{i}\in\mathfrak{R}^{b\times{\frac{h}{t}\times{\frac{w}{t}}}\times{3}}, i= 4,5,\ldots,l$. Then a concatenation layer combines the compressed features $\bm{c}_1\ldots\bm{c}_{l}$ to $\bm{r}\in\mathfrak{R}^{{b\times{\frac{h}{t}}\times{\frac{w}{t}}\times{
e}}},e= 1, 2, \ldots,3\times{l}-4$. Note that the downsampling factor $t$ determines the code length $\bm{c}_1\ldots\bm{c}_{l}$. Afterward, a power-normalization layer is applied to generate $k$ complex transmitted symbols 
$\bm{x}$ and regulate the transmitting power of these symbols to be under the given value where $\bm{x}\in\mathfrak{C}^{{b\times k}}$ and $k={\frac{h}{t}}\times{\frac{w}{t}}\times{e}$. Note that the number of \textit{l} is self-adaption according to the channel environment. The compression ratio ($k/n=\frac{3\times{t^2}}{g}, g=1, \ldots, 3\times{l}-4$) is determined by the \textit{l}. The high-level semantic features are the boost information for reconstruction, when $e=1$, the output of the Joint Semantic-channel Encoder is $\bm{c}_3\in\mathfrak{R}^{b\times{\frac{h}{t}\times{\frac{w}{t}}}\times{1}}$.

The reshape layer at the receiver first reshaped the received signals into the size of ${b\times{\frac{h}{t}}\times{\frac{w}{t}}\times{(3\times{l}-4})}$ and then separates different-level semantic features $\bm{q}_i$ from the received signals, $i=1, \ldots, l$. More specifically, $\bm{q}_1$ takes the first element of the last dimension of the reshaped symbols such that $\bm{q}_1 \in \mathfrak{R}^{b\times{\frac{h}{t}\times{\frac{w}{t}}}\times{2}}$, where the last dimension is doubled as we concatenate the real and imaginary parts of the received symbols. Similarly, $\bm{q}_i$ takes the $3i-1$th to $3i+1$th elements of the last dimension of the reshaped symbols such that $\bm{q}_i \in \mathfrak{R}^{b\times{\frac{h}{t}\times{\frac{w}{t}}}\times{6}}$. Each of $\bm{q}_i$, $i=1,\ldots,l$ is then input to a Decoder Module, the structure of which is shown in Fig. 3. The Decoder Modules have the same kernel size (\textit{m}=3). Then each of the Decoder Modules outputs the reconstructed semantic features $\bm{d_i}\in\mathfrak{R}^{b\times{{\frac{2h}{t}}\times{\frac{2w}{t}}\times{o}}}$, $i= 1,\ldots,l$, where $o$ is the output channel. Note that the output channel of $d_1$ and $d_2$ is $\frac{o}{2}$.

\subsection{Image Reconstruction Module}
The output features $\bm{d_i}$ are first combined by a concatenation layer into $\bm{v}\in\mathfrak{R}^{b\times{{\frac{2h}{t}}\times{\frac{2w}{t}}\times{(o\cdot{(l-1)})}}}$. A Multi-features fusion modSzhongule, as shown in Fig. 3, is used for the image reconstruction. Specifically, this module consists of several transpose convolutional layers with PReLU activation (sigmoid nonlinearity in the last layer), which outputs ${\hat{\bm{S}}}$ of the same size as the input image $\bm{S}$. Finally, a denormalization layer multiplies each value of ${\hat{\bm{S}}}$ by 255 to generate the pixel value within the RGB range.

\section{Numerical results}


We evaluate the proposed system using MSCOCO\cite{lin2014microsoft} and ADE20K\cite{zhou2017scene} datasets. The MSCOCO dataset contains 123287 images (82783 for training and 40504 for validation, respectively). Each image is associated with five different captions. The ADE20K dataset contains 27574 images of 150 semantic labels with at least 512 pixels in height and width.  Note that each image is cropped into a fixed size of $h = 128$ and $w = 128$ for training. 
We use the MSCOCO dataset with the text labels to train the semantic feature module, CaptionNet \cite{he2016deep,yu2019review}, and the ADE20K dataset with segmentation labels to train the rest of the system while the semantic feature module is frozen. After training, We also test the proposed system with the commonly-used image dataset Kodak \cite{kodak1993kodak}, which contains a total of 24 images of a fixed size 768 × 512.

The proposed model was implemented in Pytorch \cite{paszke2019pytorch} and optimized using the Adam algorithm \cite{kingma2014adam}.
We set the learning rate to be 0.0001, the batch size to be 32, and the downsampling factor $t=8$, respectively. 
We adopt the existing DL-based method\cite{bourtsoulatze2019deep} and \cite{yan2021deep}, referred to as DeepJSCC and Aided Deep-JSCC, and traditional separation-based digital transmission schemes (JPEG) as the baseline to compare with, and use the PSNR and SSIM metrics to evaluate performance \cite{wang2004image}. We note that Aided Deep-JSCC is a distributed semantic communication scheme for wireless sensor networks, where each sensor observes and encodes a common image to send over a wireless channel to the receiver independently, and the receiver fuses the received messages from different sensors to reconstruct the original images. This scheme is similar to our proposed scheme that multiple versions of semantic information are extracted from the original images and sent over to the receiver. We would like to note that the transmission rate must be lower the channel capacity according to Shannon’s separation theorem. The prerequisite of the source information transmitted over the physical channel is $R\leq\frac{k}{n}\rm{log_2}(1+\rm{SNR})$, where $R$ is the source information transmission rate. We can get the upper bound value $R_{max}$ which is the maximum transmission rate of the source information reliably over the physical channel. Moreover, the traditional image compression schemes have a minimum compression bitrate $R_{min}$, which is the bound of the receiver to reconstruct the original image. If $R_{min}>R_{max}$, the receiver cannot reconstruct the input image. Where $\rm{log_2}(1+\rm{SNR})$ is the channel capacity, $k$ is the number of channel dimensions, and $n$ is the number of image dimensions \cite{bourtsoulatze2019deep}, respectively. Additionally, the JPEG scheme cannot improve the reconstruction image quality from better channel conditions; once the channel SNR is under the threshold, the JPEG scheme cannot reconstruct the image but a random image instead. We assume the transmission operates at channel capacity, which means that any explicit practical channel coding and modulation schemes have worse performance than this scheme for JPEG approach.

\newcommand{\mysize}{8.4cm}
\begin{figure}[htbp]
\centering
\subfigure[]{
\includegraphics[width=\mysize]{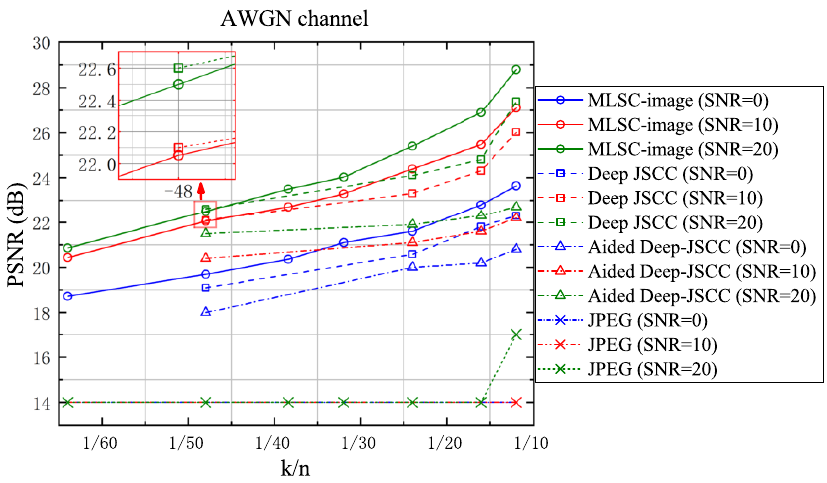}
}
\centering
\subfigure[]{
\includegraphics[width=\mysize]{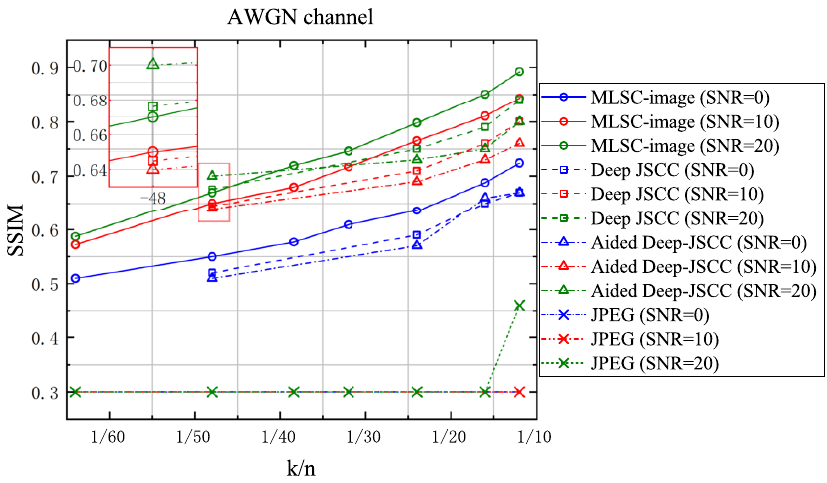}
}
\caption{Performance comparison of different approaches over an AWGN channel with different compression ratios ($k/n$) and SNR values in terms of (a) PSNR and (b) SSIM.}
\end{figure}

Fig. 4(a) shows the performance comparison with different values of $k/n$ with various SNR values in terms of PSNR, which shows that MLSC-image outperforms DeepJSCC, Aided Deep-JSCC, and traditional image compression method JPEG. We also observe that DL-based image communication systems do not suffer from unexpected performance degradation from the ``cliff effect''. That is, if channel condition is worse than a certain threshold, the receiver can not recover the transmitted images at all. It can be seen that the digital transmission schemes would break down in poor channel conditions (SNR$<$10dB) and a low compression ratio regime($k/n<1/10$), while the DL-based systems still work. We note that the DeepJSCC scheme yields a higher PSNR score than the MLSC-image in the low compression ratio (e.g., 1/48) regimes since the transmitter has plenty of bandwidth to convey more detailed information of the image. However, the proposed MLSC-image outperforms the DeepJSCC scheme when $k/n$ is bigger. Moreover, the MLSC-image has better robustness in a low compression rate with bad physical channel conditions. This is due to the fact that the proposed method uses more reasonable features extractors and transmits more significant high-level semantic information with limited bandwidth.

We present the performance comparison of different approaches in terms of SSIM in Fig. 4(b) under the same setting. The SSIM score reflects the similarity between original and reconstructed images from structural aspects. We observe a similar trend that the proposed MLSC-image remarkably outperforms the other counterparts in the low $k/n$ regime with bad channel conditions (e.g., SNR=0). However, the loss function of the Aided Deep-JSCC system with SSIM and MSE induces the scheme to have better performance on SSIM metrics. 
We note that Aided Deep-JSCC results in the worst reconstruction quality at high $k/n$ regimes. This is due to that each sensor in their setting encodes the image independent which introduces redundancy in the extracted semantic information across different encoders. Whereas, the proposed scheme extracts the multi-level semantic information in a joint manner, which enforces the semantic information obtained by different encoders to be different. The MLSC-image relatively improves the score of SSIM with the bad physical channel (e.g., SNR$\leq$10) on different $k/n$ regimes, which further demonstrates the advantages of the proposed scheme under the limited bandwidth with terrible channel conditions.

We then evaluate the robustness of the MLSC-image scheme under different channel conditions. Specifically, we train the proposed model under channels with a specific SNR value, while the model makes testing with various SNR values, and the results show in Fig. 5. Note that during the $\rm{SNR_{test}}<\rm{SNR_{train}}$ situation, the MLSC-image system does not suffer from the ``cliff effect'', which often is observed by the digital transmission scheme. 
Instead, the MLSC-image system shows a smooth decline in performance with the decrease of the SNR value. 
It exhibits an apparent tradeoff between the compression and the robustness of the proposed method. Where if the model is trained with high SNR, the PSNR performance of the system is mainly determined by the bandwidth compression ratio. More significantly, the system is robust to different channel conditions, such that the scheme does not sustain the ``cliff effect'' that appeared in the traditional communication system.

\begin{figure}[htbp]
\centering
\includegraphics[scale=0.8]{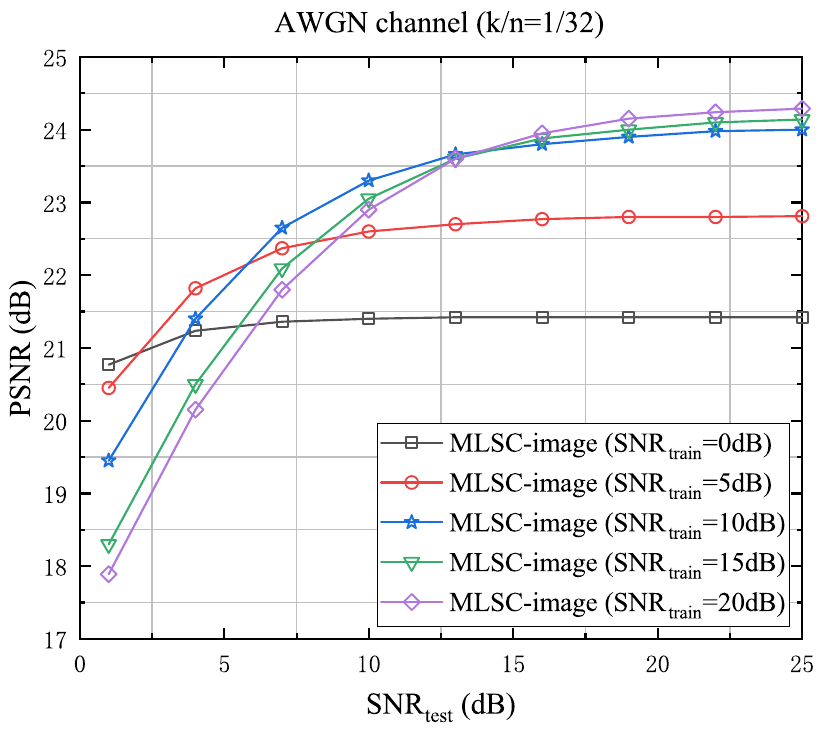}
\caption{PSNR performance of the MLSC-image algorithm over an AWGN channel with training on a specific channel SNR value and testing on the varying SNR values.}
\label{fig:5}
\end{figure}

Fig. 6 compares the PSNR performance of the alternative MLSC-image architectures without the semantic feature module or segmentation feature module. The results show that both the two modules improve the performance of the proposed system, which is more significant in the low SNR regime. It further validates the benefits of high-level text and segmentation semantics in the compression and transmission of images. We also note that the performance degradation by the model without the semantic feature module is more noticeable than that by the model without the segmentation feature module, which indicates that the semantic-related feature is of higher semantic importance.


\begin{figure}[htbp]
\centering
\includegraphics[scale=0.75]{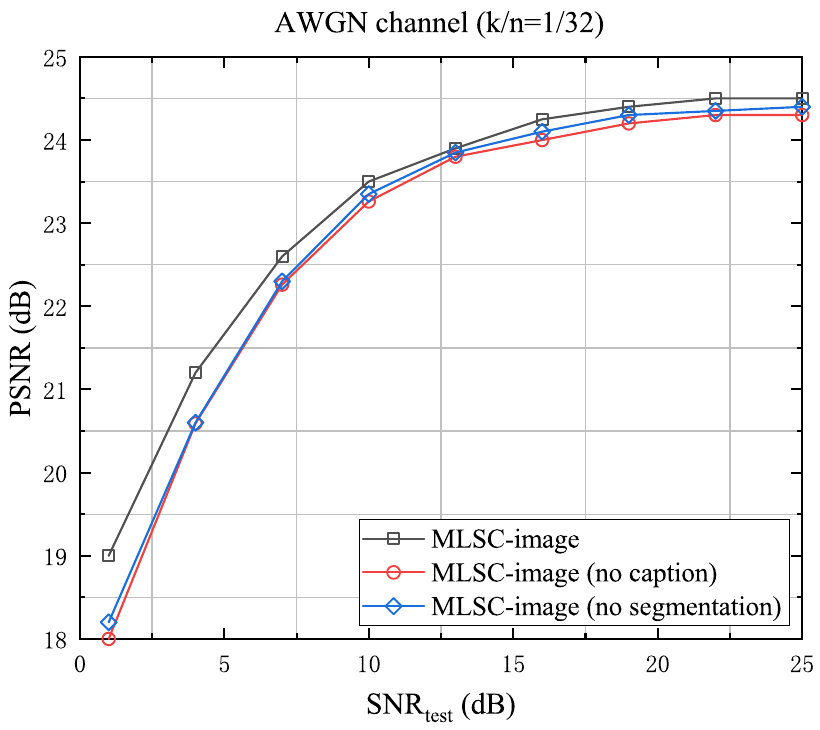}
\caption{Performance of alternative MLSC-image architectures over an AWGN channel with riding caption or image segmentation.}
\label{fig:6}
\end{figure}

We further evaluate the trained MLSC-image and benchmark approaches on the commonly-used Kodak dataset and present the comparison results in terms of PSNR with $k/n$ set to be $1/16$ under AWGN channels. As it is shown, our scheme outperforms other DL-based systems, which shows the advantages of the proposed method by exploiting the multi-level semantic information. We also note that the DL-based system exhibits smooth performance degradation when the channel condition gets worse.

\begin{figure}[htbp]
\centering
\includegraphics[scale=0.75]{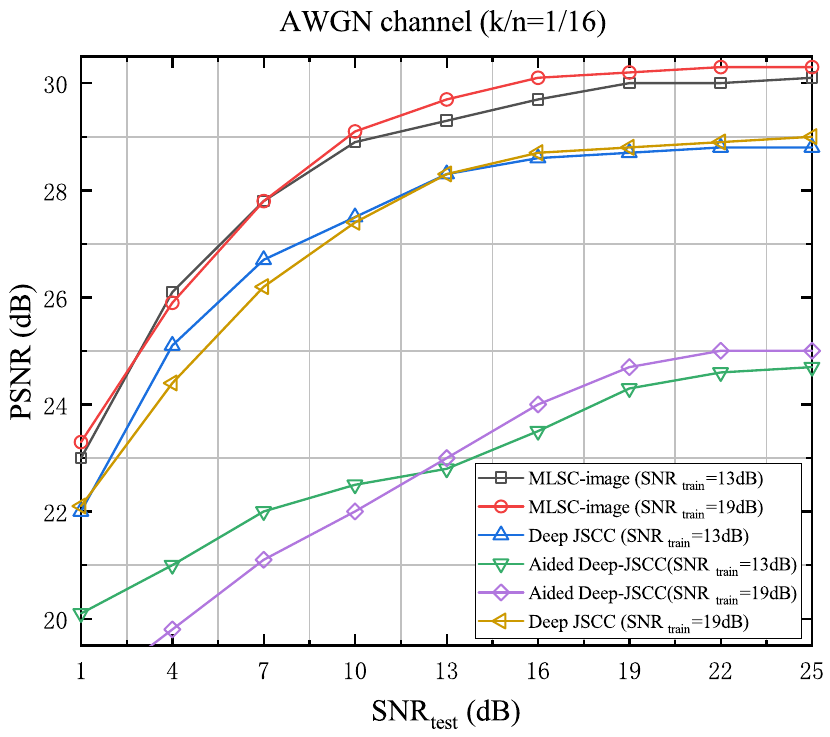}
\caption{Performance comparison in PSNR on Kodak dataset.}
\label{fig:7}
\end{figure}

\section{Conclusion}
In this paper, we proposed a novel DL-enabled semantic communication system for wireless image transmission, named MLSC-image. We jointly design the transmitter and receiver to reconstruct the source information, which achieves a remarkable performance compared to DeepJSCC and traditional separation-based digital transmission schemes. Additionally, the multi-level semantic feature extractor is used for extracting abundant features of the image with different forms. Simulation results demonstrated that the MLSC-image performance is worse than DeepJSCC due to the abundant bandwidth transmitting more detailed information in the high compression ratio situation with better channel conditions. However, the MLSC-image shows the importance of the construction features in the different compression ratio situations, especially in the low SNR regime.

\bibliographystyle{ieeetr}

\bibliography{conference}

\end{document}